\documentclass{elsart} \usepackage{graphicx} \usepackage{amssymb}
\usepackage{amsmath}

\begin{document}
\newcommand{\aff}[2]{Dipartimento di Fisica dell'Universit\`a  #1 e
  Sezione INFN, #2, Italy.}  \newcommand{\affuni}[2]{Dipartimento di
  Fisica dell'Universit\`a  #1, #2, Italy.}
\newcommand{\affinfn}[2]{INFN Sezione di #1, #2, Italy.}
\newcommand{\affinfnm}[2]{INFN Sezione di #2, #2, Italy.}
\newcommand{\affinfnn}[2]{INFN Sezione di #1, #2, Italy.}
\newcommand{\affd}[1]{Dipartimento di Fisica dell'Universit\`a  e
  Sezione INFN, #1, Italy.}

\newcommand{\dafne}	{DA$\Phi$NE }

\newcommand{\phippp}	{\phi \rightarrow \pi^+ \pi^- \pi^0}
\newcommand{\phietag}	{\phi \rightarrow \eta \gamma}

\newcommand{\pp}	{\pi^+ \pi^- } \newcommand{\ee}	{e^+ e^- }
\newcommand{\etap}{\eta^{\prime}} \newcommand{\etapppg}{\eta^{\prime}
  \rightarrow \pi^+ \pi^- \gamma} \newcommand{\etappg}{\eta
  \rightarrow \pi^+ \pi^- \gamma} \newcommand{\etappp}	{\eta
  \rightarrow \pi^+ \pi^- \pi^0}

\newcommand{\bharad}	{e^+ e^- \to e^+ e^- (\gamma)}
\newcommand{\eephietag}	{e^+ e^- \to \phi \to \eta \gamma}

\begin{frontmatter}

\title{Measurement of \mathversion{bold}$\Gamma(\eta \to
  \pi^+\pi^-\gamma)/\Gamma(\eta \to
  \pi^+\pi^-\pi^0)$\mathversion{normal} with KLOE experiment}

\collab{The KLOE Collaboration} \author[Na,infnNa]{F.~Ambrosino},
\author[Frascati]{A.~Antonelli}, \author[Frascati]{M.~Antonelli},
\author[Roma2,infnRoma2]{F.~Archilli}, \author[Cracow]{I.~Balwierz},
\author[Frascati]{G.~Bencivenni}, \author[Roma1,infnRoma1]{C.~Bini},
\author[Frascati]{C.~Bloise}, \author[Roma3,infnRoma3]{S.~Bocchetta},
\author[Frascati]{F.~Bossi}, \author[infnRoma3]{P.~Branchini},
\author[Frascati]{G.~Capon}, \author[Frascati]{T.~Capussela},
\author[Roma3,infnRoma3]{F.~Ceradini},
\author[Frascati]{P.~Ciambrone}, \author[Frascati]{E.~Czerwi\'nski},
\author[Frascati]{E.~De~Lucia},
\author[Roma1,infnRoma1]{A.~De~Santis},
\author[Frascati]{P.~De~Simone},
\author[Roma1,infnRoma1]{G.~De~Zorzi}, \author[Mainz]{A.~Denig},
\author[Roma1,infnRoma1]{A.~Di~Domenico},
\author[infnNa]{C.~Di~Donato\corauthref{cor}},
\ead{camilla.didonato@na.infn.it}
\author[Roma3,infnRoma3]{B.~Di~Micco\corauthref{cor}},
\ead{dimicco@fis.uniroma3.it} \author[Frascati]{M.~Dreucci},
\author[Frascati]{G.~Felici}, \author[Roma1,infnRoma1]{S.~Fiore},
\author[Roma1,infnRoma1]{P.~Franzini}, \author[Frascati]{C.~Gatti},
\author[Roma1,infnRoma1]{P.~Gauzzi},
\author[Frascati]{S.~Giovannella}, \author[infnRoma3]{E.~Graziani},
\author[Uppsala]{M.~Jacewicz\corauthref{cor}},
\ead{marek.jacewicz@physics.uu.se} \corauth[cor]{Corresponding
  author.}  \author[Frascati,StonyBrook]{J.~Lee-Franzini},
\author[Moscow]{M.~Martemianov},
\author[Frascati,Energ,Marconi]{M.~Martini},
\author[Na,infnNa]{P.~Massarotti}, \author[Na,infnNa]{S.~Meola},
\author[Frascati]{S.~Miscetti}, \author[Frascati]{G.~Morello},
\author[Frascati]{M.~Moulson}, \author[Mainz]{S.~M\"uller},
\author[Na,infnNa]{M.~Napolitano},
\author[Roma3,infnRoma3]{F.~Nguyen}, \author[Frascati]{M.~Palutan},
\author[infnRoma3]{A.~Passeri}, \author[Frascati,Energ]{V.~Patera},
\author[Roma3,infnRoma3]{I.~Prado~Longhi},
\author[Frascati]{P.~Santangelo}, \author[Frascati]{B.~Sciascia},
\author[Cracow]{M.~Silarski}, \author[Frascati]{T.~Spadaro},
\author[Roma3,infnRoma3]{C.~Taccini}, \author[infnRoma3]{L.~Tortora},
\author[Frascati]{G.~Venanzoni},
\author[Frascati,Energ,CERN]{R.~Versaci},
\author[Frascati,Beijing]{G.~Xu}, \author[Cracow]{J.~Zdebik}

\collab{and, as members of the KLOE-2 collaboration:}
\author[Frascati]{D.~Babusci}, \author[Roma2,infnRoma2]{D.~Badoni},
\author[infnRoma1]{V.~Bocci}, \author[Roma3,infnRoma3]{A.~Budano},
\author[Moscow]{S.~A.~Bulychjev}, \author[Frascati]{P.~Campana},
\author[Frascati]{E.~Dan\'e}, \author[INFNBari]{G.~De~Robertis},
\author[Frascati]{D.~Domenici}, \author[Bari,INFNBari]{O.~Erriquez},
\author[Bari,INFNBari]{G.~Fanizzi},
\author[Roma2,infnRoma2]{F.~Gonnella},
\author[Frascati]{F.~Happacher}, \author[Uppsala]{B.~H\"oistad},
\author[Energ,Frascati]{E.~Iarocci}, \author[Uppsala]{T.~Johansson},
\author[Moscow]{V.~Kulikov}, \author[Uppsala]{A.~Kupsc},
\author[INFNBari]{F.~Loddo}, \author[Moscow]{M.~Matsyuk},
\author[Roma2,infnRoma2]{R.~Messi}, \author[infnRoma2]{D.~Moricciani},
\author[Cracow]{P.~Moskal}, \author[INFNBari]{A.~Ranieri},
\author[Frascati]{I.~Sarra},
\author[Calabria,INFNCalabria]{M.~Schioppa},
\author[Energ,Frascati]{A.~Sciubba}, \author[Warsaw]{W.~Wi\'slicki},
\author[Uppsala]{M.~Wolke}


\address[Frascati]{Laboratori Nazionali di Frascati dell'INFN,
  Frascati, Italy.}  \address[Cracow]{Institute of Physics,
  Jagiellonian University, Krakow, Poland.}  \address[Mainz]{Institut
  f\"ur Kernphysik, Johannes Gutenberg - Universit\"at Mainz,
  Germany.}  \address[Na]{Dipartimento di Scienze Fisiche
  dell'Universit\`a  ``Federico II'', Napoli, Italy}
\address[infnNa]{INFN Sezione di Napoli, Napoli, Italy}
\address[Energ]{Dipartimento di Scienze di Base ed Applicate per
  l'Ingegneria dell'Universit\`a ``Sapienza'', Roma, Italy.}
\address[Roma1]{\affuni{``Sapienza''}{Roma}}
\address[infnRoma1]{\affinfnm{``Sapienza''}{Roma}}
\address[Roma2]{\affuni{``Tor Vergata''}{Roma}}
\address[infnRoma2]{\affinfnn{Roma Tor Vergata}{Roma}}
\address[Roma3]{\affuni{``Roma Tre''}{Roma}}
\address[infnRoma3]{\affinfnn{Roma Tre}{Roma}}
\address[StonyBrook]{Physics Department, State University of New York
  at Stony Brook, USA.}  \address[Beijing]{Institute of High Energy
  Physics of Academica Sinica,  Beijing, China.}
\address[Moscow]{Institute for Theoretical  and Experimental Physics,
  Moscow, Russia.}  \address[Marconi]{Present Address: Dipartimento di
  Scienze e Tecnologie Applicate,  Universit\`a Guglielmo Marconi,
  Roma, Italy.}  \address[CERN]{Present Address: CERN, CH-1211 Geneva
  23, Switzerland.}
\begin{center}
and
\end{center}
\address[Bari]{\affuni{di Bari}{Bari}}
\address[INFNBari]{\affinfn{Bari}{Bari}}
\address[Calabria]{\affuni{della Calabria}{Cosenza}}
\address[INFNCalabria]{INFN Gruppo collegato di Cosenza, Cosenza,
  Italy.}  \address[Uppsala]{Department of Nuclear and Particle
  Physics, Uppsala Univeristy,Uppsala, Sweden.}
\address[Warsaw]{A. Soltan Institute for Nuclear Studies, Warsaw,
  Poland.}

\begin{abstract}
We report the measurement of the ratio $\Gamma(\eta \to
\pi^+\pi^-\gamma)/\Gamma(\eta \to \pi^+\pi^-\pi^0)$ analyzing a large
sample of $\phi \to \eta \gamma$ decays recorded with the KLOE
experiment at the DA$\Phi$NE $e^+ e^-$ collider, corresponding to an
integrated luminosity of 558 pb$^{-1}$. The $\eta \to
\pi^+\pi^-\gamma$ process is supposed to proceed both via a resonant
contribution, mediated by the $\rho$ meson, and a non resonant direct
term, connected to the box anomaly. The presence of the direct term
affects the partial width value.  Our result $R_{\eta}=\Gamma(\eta \to
\pi^+ \pi^- \gamma)/\Gamma(\eta \to \pi^+ \pi^- \pi^0)= 0.1838\pm
0.0005_{stat} \pm 0.0030_{syst}$ is in agreement  with a recent CLEO
measurement, which differs by more 3 $\sigma$ from the average of
previous results.
\end{abstract}

\begin{keyword}
$e^{+}e^{-}$ collisions \sep $\eta$ decays

\end{keyword}
\end{frontmatter}

\section{Introduction}
\label{sec:introduction}
The Chiral Perturbation Theory (ChPT) provides accurate description of
interactions and decays of light mesons \cite{GasserLut}. The decays
$\etappg$ and $\etapppg$ are expected to get contribution from the
anomaly accounted for by the Wess Zumino Witten (WZW) term into the
ChPT Lagrangian \cite{Ben2003}. Those anomalous processes are referred 
to as box anomalies which proceed through a 
vector meson resonant contribution, described by Vector Meson Dominance (VMD).
According to effective theory \cite{Ben2003} the contribution of the
direct term should be present together with VMD. In case of $\etappg$ the
$\rho$ contribution is not dominant, this makes the partial width
sensitive to the presence of the direct term, while in case of $\etapppg$
the partial width is dominated by the resonance and the direct
term effect should be visible in the  dipion invariant mass
distribution. The present world average of the $\etappg$ partial width,
$\Gamma(\etappg) = (60 \pm 4)$ eV \cite{PDG10}, provides strong evidence 
in favour of the box anomaly, compared with value expected with and without 
the direct term, respectively $(56.3 \pm 1.7)$ eV and $(100.9 \pm 2.8)$ eV
\cite{Ben2003}. Recently CLEO \cite{Lopez07} has measured the ratio
$R_{\eta}=\Gamma(\eta \to \pi^+ \pi^- \gamma)/\Gamma(\eta \to \pi^+
\pi^- \pi^0)= 0.175\pm 0.007_{stat} \pm 0.006_{syst}$, which differs by more 
than $3\sigma$ from the average result of previous measurement
\cite{Gormley,Layter}, $R_{\eta}= 0.207\pm 0.004$ \cite{PDG06}. We present a new measurement with the highest 
statistics and the smallest systematic error ever achieved.
\section{The KLOE detector at DA$\Phi$NE}
\label{sec:detector}
The KLOE experiment operates at the Frascati $\phi$-factory, 
DA$\Phi$NE, an $e^+e^-$ collider running at a center of mass energy 
of $\sim 1020$~MeV, the mass of the $\phi$ meson. 
The detector consists of a large cylindrical Drift Chamber (DC),
surrounded by a lead-scintillating fiber electromagnetic calorimeter and
a superconducting coil around the EMC provides a 0.52~T field.
The drift chamber~\cite{DCH}, 4~m in diameter and 3.3~m long, has 12,582
all-stereo tungsten sense wires and 37,746 aluminum field wires. 
The chamber shell is made of carbon fiber-epoxy composite with an
internal wall of 1.1 mm thickness, the gas used is a 90\% helium, 
10\% isobutane mixture. 
The spatial resolutions are $\sigma_{xy} \sim 150\ \mu$m and 
$\sigma_z \sim$~2 mm and the momentum resolution is 
$\sigma(p_{\perp})/p_{\perp}\approx 0.4\%$.
The calorimeter~\cite{EMC} consists of a barrel and two endcaps, for a
total of 88 modules, and covers 98\% of the solid angle. 
The modules are read out at both ends by photomultipliers, both in
amplitude and time. 
The readout granularity is $\sim$\,(4.4 $\times$ 4.4)~cm$^2$, for a total
of 2440 cells arranged in five layers. 
The energy deposits are obtained from the signal amplitude while the
arrival times and the particles positions are obtained from the time
differences. 
Cells close in time and space are grouped into calorimeter clusters and
the cluster energy $E$ is the sum of the cell energies.
The cluster time $T$ and position $\vec{R}$ are energy-weighted averages. 
Energy and time resolutions are $\sigma_E/E = 5.7\%/\sqrt{E\ {\rm(GeV)}}$ 
and  $\sigma_t = 57\ {\rm ps}/\sqrt{E\ {\rm(GeV)}} \oplus100\ {\rm ps}$, 
respectively.
The trigger \cite{TRG} uses both calorimeter and chamber information.
Data are then analyzed by an event classification filter \cite{NIMOffline},
which selects and streams various categories of events in different
output files.

\section{Event selection}
\label{sec:eventselection}
The analysis has been performed using 558 pb$^{-1}$ from the 2004-2005
data set at $\sqrt{s} \simeq 1.02$ GeV.  Monte Carlo (MC) events are
used to simulate the signal and the background. The signal is
generated according to the matrix element quoted in \cite{Gormley}.
All MC productions account for run by run variations of the main data-taking
parameters such as background conditions, detector response and beam
configuration.\\ The final state under study is
$\pi^+\pi^-\gamma\gamma$, since at KLOE, the $\eta$ mesons are
produced together with a monochromatic recoil photon ($E_{\gamma} =
363$ MeV) through the radiative decay $\phietag$.  In the considered
data sample about $\simeq 25 \times 10^{6}\ \eta$'s are produced.  The
main background comes from $\phi \to \pi^+\pi^-\pi^0, \pi^0 \to \gamma
\gamma$ decaying to the same final state. Other backgrounds are $\phi
\to \eta \gamma\to \pi^+\pi^-\pi^0 \to \pi^+ \pi^- 3\gamma$
with one photon lost, and $\phi \to \eta \gamma, \eta \to e^+
e^-\gamma$ when both electrons are mis-identified as pions.

As first step of the analysis, a preselection is performed, requiring
at least two tracks with opposite charge pointing to the interaction
point (IP) and at least two neutral clusters in time (not associated to any
track), having energy $E_{cl} \ge 10$ MeV and polar angle in the range
$(23^\circ -157^\circ)$.  Tracks are sorted according to the distance
of the point of closest approach from the IP. The first two tracks
are selected.

\begin{figure}[htb]
\centering
\includegraphics*[width=0.75\textwidth]{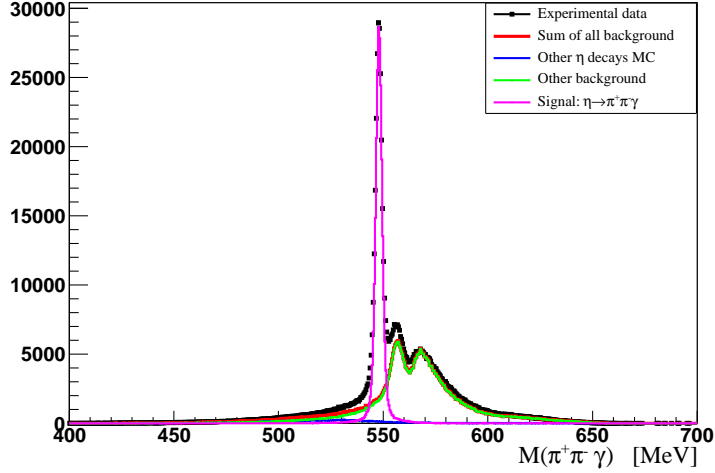}
\caption{The $\pi^+ \pi^- \gamma_{\eta}$ invariant mass distribution:
Data-MC comparison. Dots are data, Magenta is MC signal $\etappg$, 
Red is all MC background contribution}
\label{pppresults1}
\end{figure}

We require the most energetic cluster ($\gamma_{\phi}$) to have
$E_{cl}>250$ MeV and we identify it as the photon recoiling against
the $\eta$ in the $\phi \to \eta \gamma$ decay. Moreover we ask for
$\gamma_{\phi}$ inside the calorimeter barrel
($55^{\circ}-125^{\circ}$), to avoid effects of cluster merging
between barrel and end-caps of the calorimeter. Other cuts are imposed
to clean up the sample; cut on cluster-track collinearity and identification 
by time of flight (TOF) are used to reject electrons. The cut
effectively  rejects Bhabha background and other processes with
electrons in the final state.
To select $\eta$ decays we exploit the $\phi \to \eta \gamma$ two body
decay kinematic  computing the $\gamma_{\phi}$ energy, using only the
$\gamma_{\phi}$ polar angle:
\[
\vec{p}_{\phi}=\vec{p}_{\eta}+\vec{p}_{\gamma} \qquad
E_{\gamma_{\phi}}=
\frac{m_{\phi}^2-m_{\eta}^2}{2(E_{\phi}-|\vec{p}_{\phi}|cos\varphi)}
\] 
where $\varphi$ is the angle between the average  $\phi$-meson
momentum measured run by run with high accuracy and
$\gamma_{\phi}$. This allows us to improve the energy measurement of
the recoil photon to $0.1\%$. We can
determine the direction of the photon from $\eta$ decay using
$\phi$ and $\pi$-mesons information:
\[
\vec{p}_{\gamma_{\eta}} = \vec{p}_{\phi}- \vec{p}_{\pi^+} -
\vec{p}_{\pi^-}-\vec{p}_{\gamma_{\phi}}
\]
The photon direction is compared with the direction of each neutral
cluster, $\Delta \varphi = \varphi^{\mathrm{clu}} -
\varphi_{\gamma_{\eta}}$. If no cluster within $\Delta \varphi <
8.5^{\circ}$ is found the event is rejected. The cluster with the
minimum value of $\Delta\varphi$ is selected for further analysis. In
order to reject the $\phi \to \pi^+ \pi^- \pi^0$ background, the angle 
between the two photons in the $\pi^0$ reference frame, 
evaluated using the $\phi$ and the $\pi$-mesons momenta, 
is calculated and rejected with an angular cut $\varphi_{\gamma\gamma}^{\pi^+\pi^-\gamma} <165^{\circ}$; in order to reduce the systematics the angle is evaluated in the transverse plane\footnote{The $\varphi$ angle of the cluster is measured with
  an angular resolution of 6 mrad using the position of the
  calorimeter cell. The polar angle is instead determined by the time
  difference of the cluster at each side of the barrel and is
  affected by larger systematics.}. Finally we select events
requiring $539.5$ MeV $< M_{\pi^+ \pi^- \gamma} < 554.5$ MeV
(fig. \ref{pppresults1}).
\subsection{Normalization Sample: \bf $\eta \to \pi^+ \pi^- \pi^0$}
\label{sec:ppp}
\begin{figure}[htb]
\centering
\includegraphics[width=0.5\textwidth]{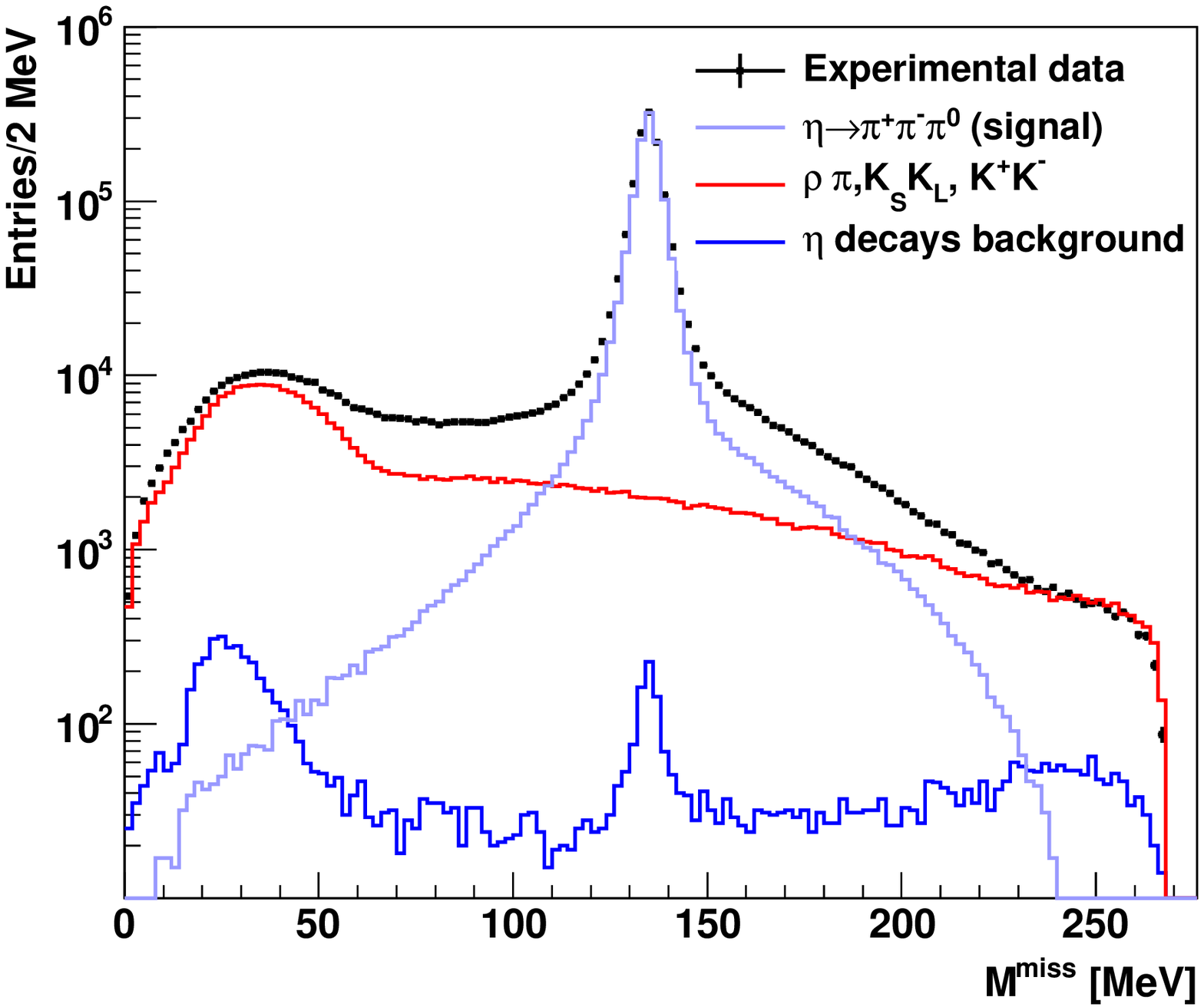}%
\includegraphics[width=0.5\textwidth]{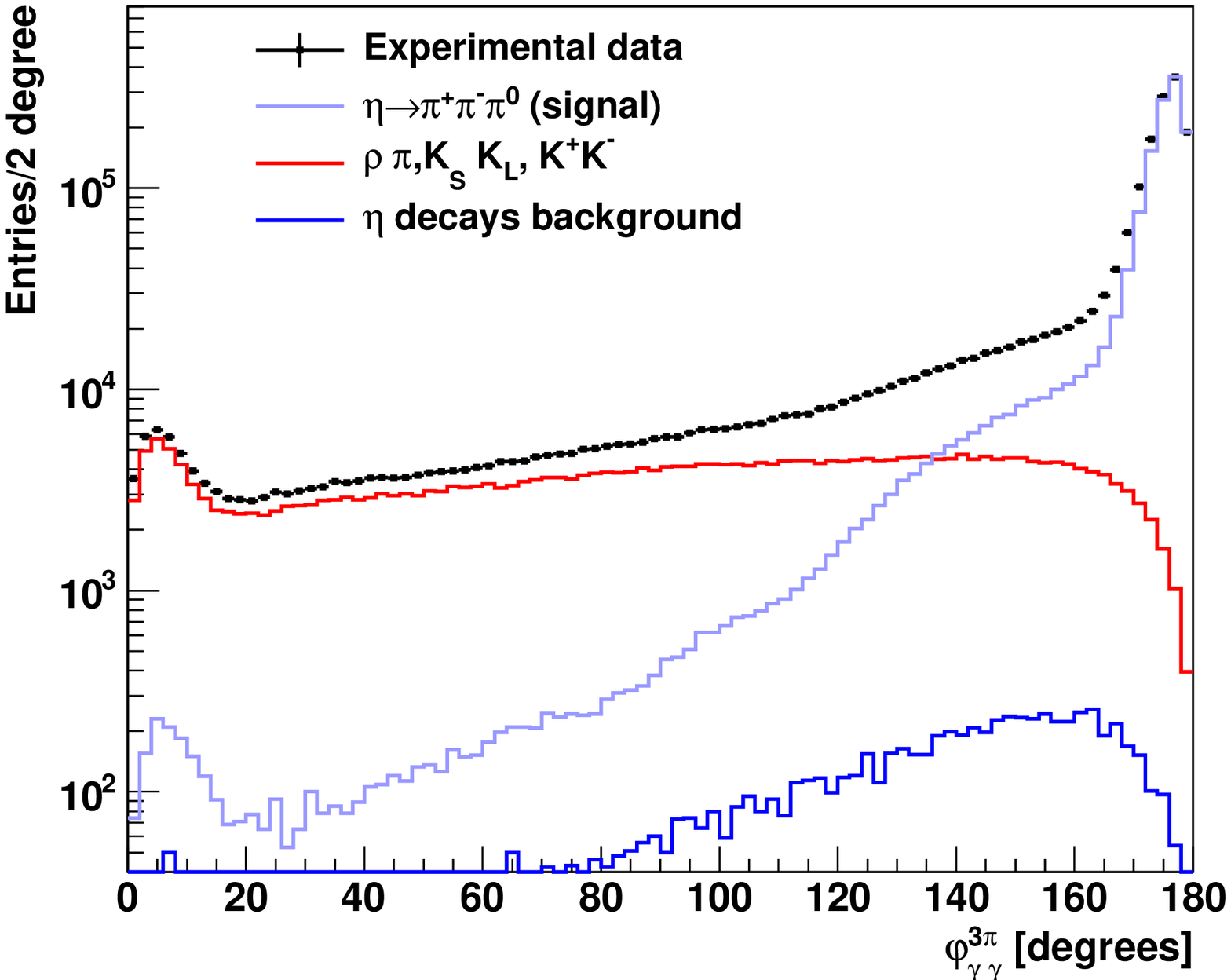}
\caption{Left - missing mass spectrum around the $\pi^0$ mass. Right -
  angle between prompt neutral clusters in the $\pi^0$ rest frame
evaluated in the transverse plane ($\varphi_{\gamma\gamma}$).}
\label{figPPP}
\end{figure}

The process $\phi \to \eta \gamma$ with $\eta \to \pi^+ \pi^- \pi^0$
represents a good control sample, due to the similar
topology. Moreover the ratio $\Gamma(\eta \to \pi^+ \pi^-
\gamma)/\Gamma(\eta \to \pi^+ \pi^-\pi^0)$ is not affected by the
uncertainties on the luminosity, the $\phi \to \eta \gamma$ partial
width and the $\phi$ production cross section cancel in the
ratio. We use the same preselection as for the $\eta \to \pi^+ \pi^-
\gamma$ signal and calculate the missing four-momentum:
\[
\mathbb{P}_{miss}=\mathbb{P}_{\phi}-\mathbb{P}_{\pi^+}-\mathbb{P}_{\pi^-}-\mathbb{P}_{\gamma_{\phi}}
\]
where the variables in the formula represent the four-momenta of the
$\phi$ meson and the products of the decays. For the  $\eta \to \pi^+
\pi^- \pi^0$ signal, the missing mass peaks at
the $\pi^0$ mass value and we select events with
$|M_{miss}-m_{\pi^0}|<15$ MeV. The remaining background is rejected
very efficiently by using an angular cut applied to the two photons
from the $\pi^0$ decay, $\varphi_{\gamma\gamma}^{3\pi}>165^{\circ}$;
the angle is evaluated in the transverse plane. 
Fig. \ref{figPPP} shows the distribution of the missing mass 
and $\varphi_{\gamma\gamma}$.
We select $N(\eta \to \pi^+\pi^-\pi^0)= 1115805 \pm 1056$, with a
selection efficiency of $\varepsilon = 0.2276\pm0.0002$ and a
background contamination of $0.65\%$. 

\section{Results}
\begin{figure}[htb]
\centering  \includegraphics*[width=0.75\textwidth]{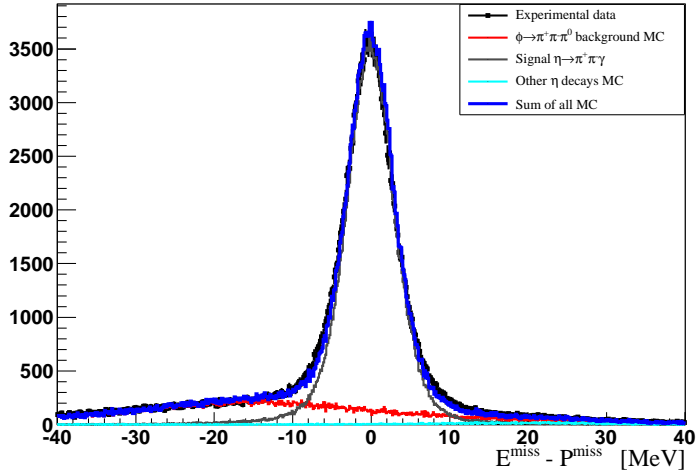}
\caption{$E_{miss}-P_{miss}$ distribution of the $\pi^+ \pi^-
  \gamma_{\phi}$ system: the fit is performed with the  background and
  signal shapes from MC in the range $|E_{miss}-P_{miss}|<10$ MeV.}
\label{EPfit}
\end{figure}

The total selection efficiency of the $\eta \to \pi^+\pi^-\gamma$
signal is $\varepsilon = 0.2131 \pm 0.0004$.  Background contribution
and the signal amount in the final sample are evaluated with a fit  to
the $E_{miss}-P_{miss}$ distribution of the $\pi^{+} \pi^{-}
\gamma_{\phi}$ system with the shapes from remaining background and
signal MC in the range $|E_{miss}-P_{miss}|<10$ MeV, fig. \ref{EPfit}. We find
$N(\eta \to \pi^+\pi^-\gamma)= 204950 \pm 450$ with a background
contamination of 10\%.
Combining our results we obtain the ratio:
\begin{equation}
R_{\eta}=\frac{\Gamma(\eta \to \pi^+ \pi^- \gamma)}{\Gamma(\eta \to
  \pi^+ \pi^- \pi^0)}= 0.1838\pm0.0005_{stat} \pm 0.0030_{syst}
\end{equation}
to be compared with world average value $\Gamma(\eta \to \pi^+ \pi^-
\gamma)/ \Gamma(\eta \to \pi^+ \pi^- \pi^0)=0.202\pm0.007$
\cite{PDG10}.
The systematic uncertainties due to analysis cuts have been evaluated
by varying the cuts on all variables and re-evaluating the
branching ratios.  The relative variation for each source of
systematic is in table \ref{tab:systematics}.  The total error is taken
as the quadratic sum of all contributions.
\begin{table}[!t]
 \begin{center}
  {\scriptsize
    \begin{tabular}{cc}
    \hline Source of uncertainty    & Relative error \\ \hline 
    $\varphi_{\gamma\gamma}^{\pi^+\pi^-\gamma} < 165^{\circ} \pm 2^{\circ} $ & $\pm0.6\%$\\       
    $\Delta \varphi > 8.5^{\circ} \pm 2^{\circ}$ & $\pm 0.4\%$\\ 
    $|M_{\pi^+ \pi^-\gamma}-M_{\eta}| < 7.5$ MeV $\pm$ 2 MeV & $\pm 0.6\%$\\ 
    $|M_{miss} - M_{\pi^0}| < 15$ MeV $\pm$ 4 MeV & $\pm 0.4\%$\\ 
    $\varphi_{\gamma\gamma}^{3\pi} >  165^{\circ} \pm 2^{\circ}$ & $\pm 0.1\%$\\ 
    $E_{min}^{\gamma} > 10$ MeV $\pm$ 2 MeV  & $\pm 0.1\%$\\ 
    $E_{clu}^{\gamma_{\phi}} > 250$ MeV $\pm$ 4 MeV  & $\pm 0.1\%$\\
Preselection & $1\%$ \\  Fit $E_{miss}-P_{miss}$ & $\pm0.6\%$ \\ \hline
Total                    & $1.6\%$ \\ \hline    
    \end{tabular}}
   \caption{Summary table of systematic uncertainties.}
  \label{tab:systematics}
  \end{center}
\end{table}

\subsection{Dipion Invariant Mass}
\label{sec:fit-invmass}
\begin{figure}[htb]
\centering
\includegraphics[width=0.7\textwidth]{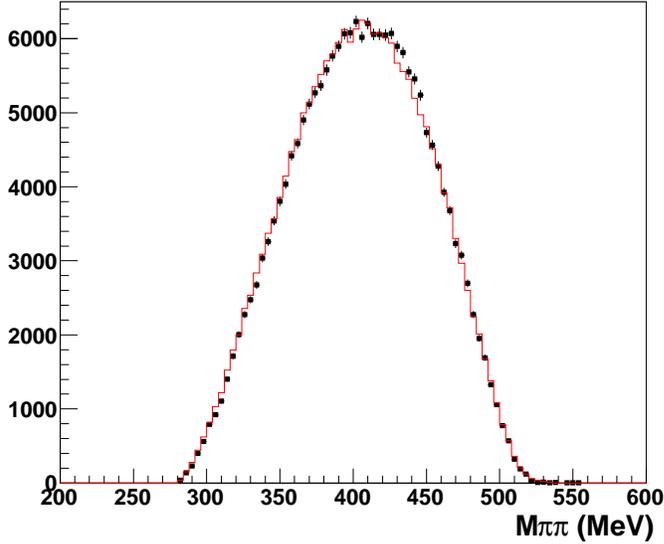}
\caption{$M_{\pi^+\pi^-}$ distribution: dots are data; histogram is the prediction from eq.\ref{eqvdm1}, corrected for acceptance and experimental
resolution}
\label{fig:Mpp}
\end{figure}

The $M_{\pi^+ \pi^-}$ dependence of decay width has been parameterized in
different approaches, in which VMD has been implemented in effective
Lagrangians \cite{Ben2003,Picciotto92}. We present a preliminary comparison
between dipion invariant mass, with the most simple approach
\cite{Picciotto92}
\begin{equation}
 \frac{d\Gamma(\eta \to\pi^+\pi^-\gamma)}{d \sqrt s} = A
 \left|1+\frac{3 m_{\rho}^2}{s-m_{\rho}^2}\right |^2
 k^3_{\gamma}q^3_{\pi}
\label{eqvdm1}
\end{equation}
where $k_{\gamma}$ is the photon momentum in the $\eta$ rest frame and
its expression is $k_{\gamma}= (m^2_{\eta}-s)/2m_{\eta}$, while
$q_{\pi}= \sqrt{s-4m_{\pi}^2}/2$ represents the pion momentum in the
dipion rest frame; $s$ is the $M_{\pi^+ \pi^-}$ invariant mass
squared.
In fig. \ref{fig:Mpp} we compare the observed $M_{\pi^+\pi^-}$ spectrum,
background subtracted, with the theoretical prediction of eq.\ref{eqvdm1}
corrected for acceptance and experimental resolution.
The agreement with data is good; fits with more complex parameterizations
are in progress.
\section{Conclusions}
\label{sec:conclusions}
Using a data sample corresponding to an integrated luminosity of 558
pb$^{-1}$, we select $204950$ $\eta \to \pi^+ \pi^- \gamma$ events and
$1115805$ $\etappp$ from the $\phi \to \eta \gamma$ decays.  The
corresponding width ratio is:
\begin{equation}
R_{\eta}= 0.1838\pm0.0005_{stat} \pm 0.0030_{syst}
\end{equation}
Our measurement is in agreement with the most recent result from CLEO
\cite{Lopez07}, which is $R_{\eta}= 0.175\pm0.007_{stat} \pm
0.006_{syst}$.\\ Combining our measurement with the world average value
$\Gamma(\eta \to \pi^+\pi^-\pi^0)= (295 \pm 16)$ eV \cite{PDG10}, we get 
$\Gamma(\eta \to \pi^+\pi^-\gamma)=(54.2\pm 0.3)$ eV,
which is in agreement with the value expected taking into account the direct
term \cite{Ben2003}, providing a strong evidence in favour of the box anomaly. 
\\The preliminary measurement of the dipion invariant mass spectrum agrees with the simplest parametrization in Hidden Symmetry model as from \cite{Picciotto92}. 

\end{document}